# Tunneling phase diagram: A machine-learning framework for multidimensional kinetic isotope effects

Xinrui Yang,[1, 2] and Zhigang Wang[1, 2 *]

[1] Institute of Atomic and Molecular Physics, Jilin University, Changchun 130012, China.

[2] Key Laboratory of Material Simulation Methods & Software of Ministry of Education, College of Physics, Jilin University, Changchun 130012, China.

[*] Correspondence email: wangzg@jlu.edu.cn (Z. W.).

## Abstract:

The kinetic isotope effect (KIE) is the conventional probe for quantum tunneling, yet its composite nature conflates tunneling with zero-point energy and classical kinetics. Here, we introduce the tunneling phase diagram—a machine-learning framework that decouples true tunneling strength by decoding the nonlinear relationship between KIE and the tunneling factor (κ). With exceptional fidelity ($R^2 > 0.98$, RMSE = 0.21), this framework reveals an anomalous high KIE–low κ spanning 300–600 K, thereby defining a paradigm for the quantitative assessment of quantum tunneling.

## Introduction:

Quantum tunneling has played an increasingly recognized role in chemical reaction dynamics [1-3] since its conceptualization in 1927 [4]. Due to the exponential sensitivity of tunneling probability to particle mass ($\propto \exp(-m^{1/2})$), isotopic substitution provides a natural experimental window into tunneling contributions. This physical intuition placed the kinetic isotope effect (KIE = kH/kD) at the center of tunneling studies since Bell's pioneering work [5-6]: a large KIE was traditionally taken as evidence of strong tunneling. Considerable efforts followed throughout the late twentieth century to establish reliable KIE-based assessment protocols [7-8], and investigations of the Swain-Schaad exponent [9-12] further enriched this pursuit in the early 2000s. The scientific view of tunneling has itself evolved, from a minor correction to classical kinetics [4] to a non-negligible factor across diverse hydrogen-transfer reactions [7-8]. This accumulated understanding increasingly calls for a shift from qualitative detection toward quantitative characterization.

A difficulty with KIE as a quantitative metric has become apparent. KIE is a composite observable whose measured values conflate tunneling with zero-point energy shifts, potential energy surface (PES) characteristics, and classical kinetic factors [15-16,19]. Notably, anomalously small KIE values have been observed in systems with clear tunneling [13-18], while large KIE values can arise from non-tunneling origins [19-20]. These observations suggest that KIE alone may not reliably gauge tunneling strength. Theoretically, the tunneling correction factor κ, defined in transition state theory as the ratio of quantum to classical rates [21-22], isolates the tunneling contribution directly and has been widely adopted as a fundamental QMT metric [23-32]. The mapping from the experimentally accessible KIE to the intrinsic κ is not straightforward: it is jointly modulated by temperature, reduced mass, and PES topology [33-38], forming a high-dimensional nonlinear relationship that defies analytical solution. Machine learning, which has recently shown encouraging success in predicting reaction rate constants [39-41], offers a promising route to decode this mapping.

In this work, we combine first-principles rate calculations with ensemble machine learning (ML) to examine KIE's reliability as a tunneling indicator and to construct a quantitative mapping from experimental observables to κ. Using density functional theory (DFT) data for the chiral inversion of four amino acids, we benchmarked eight ML algorithms and employed SHAP (SHapley Additive exPlanations) analysis to assess descriptor contributions to the KIE–κ relationship. The results show that ensemble models predict κ with substantially higher accuracy than KIE alone. The resulting multidimensional framework, which we term a "tunneling phase diagram," reveals distinct regimes across temperature, including a region of high KIE yet low κ, suggesting that the KIE–κ relationship is more complex than a simple linear correspondence. This finding offers a new perspective on the quantitative assessment of QMT.

## Methodology:

### 1. Theoretical Framework and Feature Selection

Within the framework of transition state theory (TST), the quantum reaction rate constant $k_{Tun(T)}$ is defined as the product of κ and the classical reaction rate constant $k_{Cla(T)}$:

$$k_{Tun}(T)=\kappa(T)\cdot k_{Cla}(T)$$

The classical rate constant $k_{Cla(T)}$, dominated by the Arrhenius equation, is given by:

$$k_{Cla}(T)=\frac{k_B T}{\mathrm{h}} e^{-\beta E}$$

where $k_B$ is the Boltzmann constant, h is Planck's constant, E is the activation energy, and $\beta=1/(k_B T)$. This equation indicates that the classical rate is governed solely by temperature and the barrier height, exhibiting exponential growth with increasing temperature. In contrast, quantum effects enter entirely through κ. According to Variational Transition State Theory (VTST), κ(T) can be expressed as:

$$\kappa(T)=\beta e^{\beta E^{+}} \int_{E_0}^{+\infty} P_{Tun}(E,H)\, e^{-\beta E} dE$$

$$\kappa(T)=\frac{\int_{E_0}^{+\infty} P_{Tun}(E,H)\, e^{-\beta E} dE}{\int_{E_0}^{+\infty} P_{Cla}(E,H)\, e^{-\beta E} dE}$$

Here, the analytical expression for the quantum tunneling probability $P_{Tun}$ relies on the Wentzel-Kramers-Brillouin (WKB) approximation, specifically:

$$P_{tunneling}=Exp\left[-\frac{2}{\hbar}\int_{x_1}^{x_2}\sqrt{2m(V(x)-E)}dx\right]$$

This expression captures the dependence of tunneling probability on the PES morphology (barrier height V(x), width $x_2-x_1$, and asymmetry η) as well as the reduced mass m and the energy E, underscoring its multi-parameter nature.

Through isotopic substitution (H→D), KIE is expressed as:

$$\frac{\kappa(H)}{\kappa(D)}=\frac{\int_{E_0}^{+\infty}[P_{Tun}(E,H)+P]e^{-\beta E}dE}{\int_{E_0}^{+\infty}[P_{Tun}(E,D)+P]e^{-\beta E}dE}$$

Thus, the numerical relationship between KIE and κ is governed not only by the tunneling effect itself but also by the PES topology and thermodynamic parameters, providing a multi-dimensional physical feature space for machine learning.

**Table 1**. Descriptors, symbols, and their descriptions involved in this chapter.

| Descriptor | Symbol | Description |
|---|---|---|
| Temperature | T | Directly measurable, a core thermodynamic parameter regulating reaction rates. |
| Quantum reaction rate | $k_{Tun}$ | Integrates barrier height, width, and shape, bridging theoretical calculations and experimental data. |
| Barrier asymmetry | η | Reflects stability differences between reactants and products, derived from experiments and theory. |
| Kinetic isotope effect | KIE | Experimentally measurable ratio of isotopic substitution effects on reaction rates. |

Descriptor selection was guided by two criteria: experimental measurability and theoretical completeness to capture physically meaningful relationships across experimental and theoretical domains. Among the factors influencing κ, temperature (T) was prioritized for feature space construction as it is a core, directly measurable thermodynamic parameter that governs the classical reaction rate. Parameters such as reduced mass (m) and barrier width, despite their clear physical significance in theoretical models, were excluded due to limitations in experimental resolution for direct observation. Although the barrier height (E) cannot be measured directly in experiments, it can be inversely derived via the Arrhenius equation ($k_{Cla} \propto e^{-E/(kT)}$) combined with reaction rate data; this indirect accessibility ensures its applicability within an experiment-theory correlation framework. The barrier asymmetry parameter (η) (reflecting the stability difference between reactants and products), can be inferred indirectly from spectroscopic transition-state features or PES calculations, and was therefore included.

Consequently, this study selected KIE, the quantum reaction rate ($k_{Tun}$), T, and η as the core descriptor set. We note that $k_{Tun}$ incorporates κ by definition ($k_{Tun} = \kappa \cdot k_{Cla}$); however, because $k_{Tun}$ is an experimentally accessible composite observable while κ is a derived theoretical quantity, using $k_{Tun}$ as a predictor of κ does not introduce circularity in practice: while κ can be formally extracted from $k_{Tun}/k_{Cla}$ when both rates are known from theory, experimental measurements typically provide only $k_{Tun}$ (via rate measurements) and KIE (via isotopic substitution), without direct access to $k_{Cla}$ or κ. The ML model therefore learns to invert the implicit κ-dependence embedded in the composite observable $k_{Tun}$, constrained by its covariation with KIE, T, and η across the multidimensional dataset.

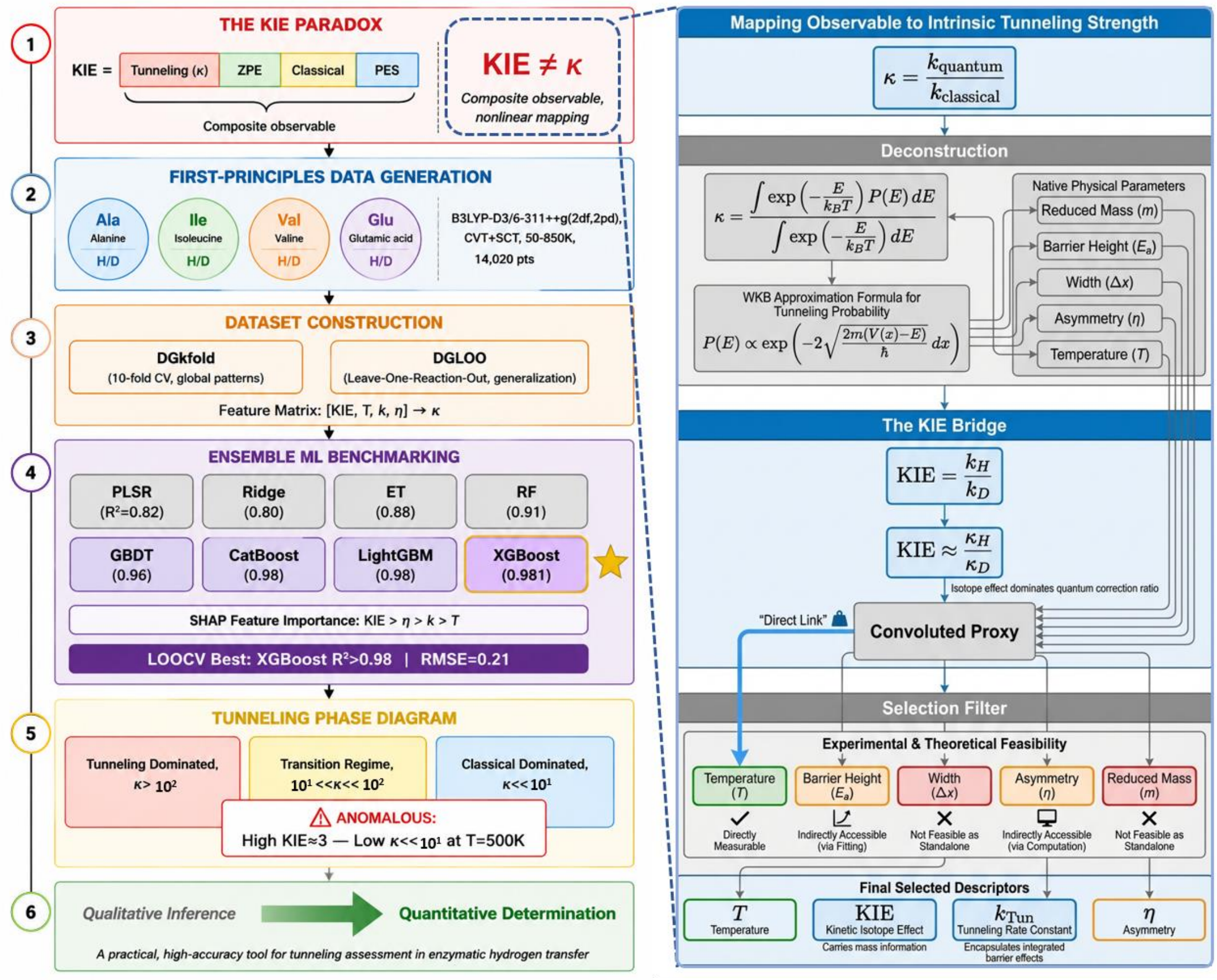

**Fig. 1.** Machine-learning framework and theoretical foundation for decoding the KIE–κ relationship. (Left) Conceptual workflow: the KIE paradox (KIE as a composite observable), first-principles data generation (DFT calculations for four amino acids with H/D substitution), ensemble ML benchmarking (eight algorithms with XGBoost optimal), and the tunneling phase diagram revealing three dynamical regimes and an anomalous high KIE–low κ region. (Right) Theoretical derivation of the descriptor set: $k_{Tun} = \kappa \cdot k_{Cla}$, with κ obtained via VTST/SCT and KIE from isotopic substitution, establishing the feature space [KIE, T, k, η] → κ.

## 2. Dataset Construction and Augmentation

The dataset was constructed using DFT computational results for the rate-limiting step of chiral inversion in four amino acid molecules (alanine, isoleucine, valine, and glutamic acid). These calculations were performed at the B3LYP-D3/6-311++g(2df,2pd) theoretical level [42] with tunneling corrections to construct the PESs. The classical rate constants $k_{Cla}$, excluding tunneling effects, were derived using canonical variational transition state theory (CVT). Eigenstate information was obtained by interfacing the Gaussrate software with Gaussian and Polyrate packages [43-45]. To reliably quantify the contribution of multidimensional tunneling effects to reaction kinetics, small-curvature tunneling (SCT) corrections were applied to obtain quantum reaction rates $k_{Tun}$ [46-47]. The temperature range was set from 50 to 1000 K, with data points sampled at 50 K intervals (20 points in total). Classical reaction rate constants and their deuterium-substituted counterparts were also calculated. The chiral inversion proceeds via a concerted proton transfer mechanism assisted by a water molecule acting as a proton shuttle, forming an amino acid–water complex.

The lower temperature bound (50 K) was selected to probe the deep-tunneling regime where quantum effects are most pronounced. While gas-phase amino acids at 50 K are metastable and may undergo condensation or structural rearrangements under experimental conditions, the computational framework employed here treats isolated molecules in the gas phase, enabling systematic exploration of the intrinsic temperature dependence of tunneling without condensed-phase complications. The primary objective is to establish the methodological validity of the KIE–κ mapping across the full range of tunneling regimes, from deep quantum to classical, rather than to predict experimental rates under cryogenic conditions.

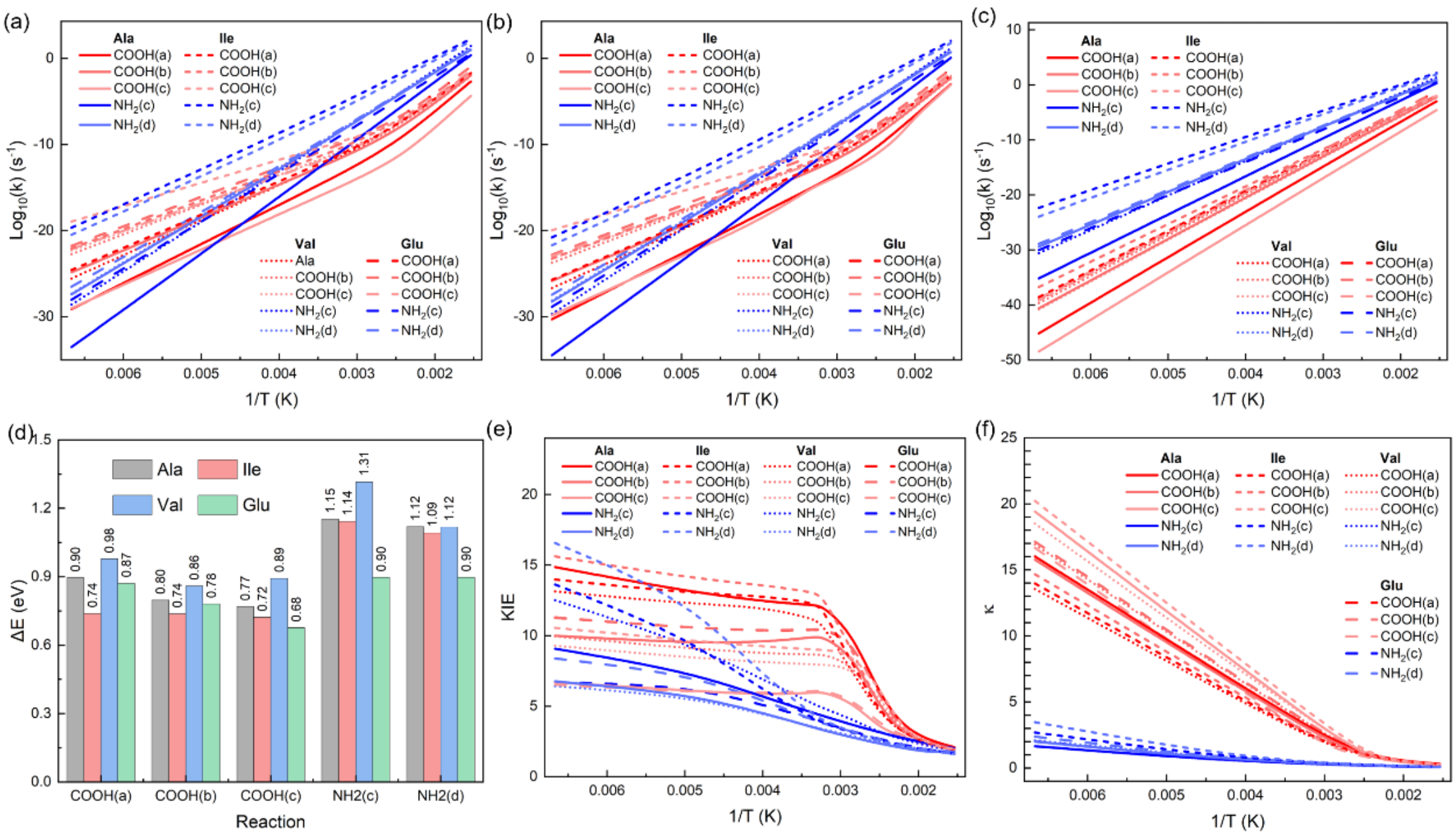

**Fig. 2.** Presentation of the reaction dataset for the concerted proton transfer reaction pathway mediated by a water bridge. (a) and (b) depict the quantum reaction rates as a function of temperature for the COOH and $NH_2$ reactions of four amino acid molecules under different hydrogen isotope conditions (H/D). (c) illustrates the classical reaction rates as a function of temperature for various reactions. (d) provides a bar chart showing the reaction asymmetry for different reactions. (e) displays the kinetic isotope effect values as a function of temperature for distinct reactions. (f) shows the relationship between the tunneling correction factors and temperature for different reactions.

As illustrated in Fig. 2(a)-(b), logarithmic quantum reaction rates (logk) for the four amino acids exhibit linear relationships with inverse temperature (1/T), where the slopes and intercepts correspond to activation energies (Ea) and pre-exponential factors (A), respectively. For instance, glutamic acid (Glu) in the $NH_2$(c) pathway shows a low activation energy (Ea=84.1 kJ/mol) and an anomalously low pre-exponential factor ($A=6.40\times10^{7}\ s^{-1}$), indicating superior reactivity attributed to the dynamic relaxation of hydrogen-bond networks in its transition state. Fig. 2(a)-(c) confirm the exponential growth of k with increasing temperature (1/T decreasing), consistent with thermally activated mechanisms. Fig. 2(d) compares the asymmetry (η) between COOH and $NH_2$ pathways, revealing higher asymmetry in $NH_2$ reactions, which suppresses quantum tunneling (Fig. 2(e)-(f)). The polar carboxyl group in COOH reactions creates steeper barriers, enhancing tunneling probabilities, while the higher asymmetry of $NH_2$ barriers suppresses quantum penetration. Furthermore, KIE and κ exhibit nonlinear temperature dependencies: KIE shows an inflection near 300 K, whereas κ monotonically decreases with temperature. These trends arise from the mass sensitivity of KIE and the direct dependence of κ on barrier penetration, modulated by PES topology and thermodynamic conditions.

To enrich the dataset, the original reaction rate data from Polyrate were expanded using a three-parameter Arrhenius equation:

$$k=AT^{m}e^{-\frac{E_a}{RT}}$$

where A is the pre-exponential factor, T is temperature, m is an additional temperature exponent, Ea is the activation energy, and R is the universal gas constant. Rate constants were regenerated at 1 K intervals to capture nonlinear temperature dependencies, a common practice in ML studies of reaction kinetics [48-51]. Parameters for $k_{Tun}$(H), $k_{Tun}$(D), and $k_{Cla}$(H) were fitted and used to extend KIE and κ values.

We acknowledge that this approach effectively interpolates between the originally computed points rather than generating independent data, which may propagate fitting errors into the augmented dataset. However, given the smooth, thermally activated nature of the Arrhenius dependence, such interpolation introduces minimal bias while significantly expanding the feature space for machine learning.

## 3. Machine Learning Models and Evaluation Framework

**Table 2.** Datasets obtained through different partitioning methods to achieve distinct training objectives.

| Dataset | Dataset | Composition | Sample size |
|---|---|---|---|
| DGLOO | Training<br>Validation | 19 reaction pathways | 38000 |
| | Test | One of the reaction pathways | 2000 |
| DGkfold | Training<br>Validation | 90% Data | 36000 |
| | Test | 10% Data | 4000 |

A tiered validation framework was established by constructing two functionally complementary dataset groups (DG), as outlined in Table 2. The first group, DGkfold, served as the foundational training architecture. It employed 10-fold cross-validation, where the raw data were systematically partitioned via a pure random split into training, validation, and test sets into approximately 90% training and 10% validation subsets per fold, with an independent 10% test set held out prior to cross-validation. This design ensures the model captures universal patterns in reaction kinetics across the global data distribution. Ten rounds of iterative training effectively balance model capacity and generalization performance, establishing a robust benchmark for subsequent evaluation.

For the model assessment phase, a more rigorous testing environment was implemented using the DGLOO dataset and Leave-One-Out Cross-Validation (LOOCV). The core mechanism of this approach involves isolating the entire temperature-dependent dataset of one specific reaction system as the independent test set. The data from the remaining n-1 reaction systems are then divided into training and validation sets via stratified random sampling at an 19:1 ratio. This leave-one-reaction-system-out strategy explicitly tests the model's predictive capability and transferability to entirely unseen reaction pathways by excluding all spatiotemporal correlation data associated with the held-out system. The stratified sampling process maintains the distribution of different reaction types.

A systematic evaluation of eight distinct machine learning algorithms was conducted for predicting quantum tunneling effects. The assessed models included: Partial Least Squares Regression (PLSR), Ridge Regression, Extra-Trees (ET), Random Forest (RF), Gradient Boosting Decision Trees (GBDT), and the ensemble frameworks CatBoost, LightGBM, and XGBoost. All models were implemented within the Python ecosystem. PLSR, Ridge, ET, RF, and GBDT were implemented using Scikit-learn, while CatBoost, LightGBM, and XGBoost were implemented via their official packages.

During the hyperparameter optimization phase, Bayesian Optimization was employed, superseding traditional grid search. This approach utilizes a Tree-structured Parzen Estimator (TPE) [52-53] to implement a sequential parameter space exploration strategy, achieving efficient optimization within limited computational resources. For instance, when optimizing XGBoost, key parameters such as the learning rate ($\eta \in [0.01, 1.0]$), maximum tree depth ($d_{max} \in \{1, 3\}$), and minimum child weight ($\gamma \in [10^{-3}, 10]$) were tuned. The results of the hyperparameter optimization are summarized in Supporting Information.

Root Mean Squared Error (RMSE) was employed as the primary loss function during model training due to its effectiveness in promoting model robustness. To comprehensively evaluate model performance, a multi-dimensional assessment system was implemented, incorporating four additional metrics alongside RMSE: Mean Absolute Error (MAE), Mean Squared Error (MSE), the coefficient of determination ($R^2$), and the Deviation. MAE reflects the average deviation of predicted values via the mean of absolute errors, offering lower sensitivity to outliers and suitability for evaluating general model performance. MSE, the precursor to RMSE, emphasizes the weight of larger errors by averaging the squared differences. $R^2$ quantifies the proportion of the variance in the target variable that is predictable from the independent variables, with values closer to 1 indicating a superior fit.

$$MAE=\frac{1}{N}\sum_{i=1}^{N}\left(k_{pred}^{(i)}-k_{obs}^{(i)}\right)$$

$$MSE=\frac{1}{N}\sum_{i=1}^{N}\left(k_{pred}^{(i)}-k_{obs}^{(i)}\right)^2$$

$$RMSE=\sqrt{\frac{1}{N}\sum_{i=1}^{N}\left(k_{pred}^{(i)}-k_{obs}^{(i)}\right)^2}$$

Here, $k_{obs}$ represents the target rate constant, $k_{pred}$ denotes the predicted rate constant, i indexes an individual data point in the dataset, and N is the total number of sampled rate constants.

$R^2$ is defined via the ratio of the residual sum of squares (SSR) to the total sum of squares (SST):

$$R^2=1-\frac{SSR}{SST}=1-\frac{\frac{1}{N}\sum_{i=1}^{N}\left(k_{pred}^{(i)}-k_{obs}^{(i)}\right)^2}{\frac{1}{N}\sum_{i=1}^{N}\left(k_{pred}^{(i)}-\bar{k}_{obs}^{(i)}\right)^2}$$

The Deviation is calculated as follows:

$$Dev=10^{x\sqrt{\frac{1}{N}\sum_{i=1}^{N}\left(k_{pred}^{(i)}-k_{obs}^{(i)}\right)^2}}-1$$

## Results and Discussion:

### 1. Comparative Model Performance and Benchmark Evaluation

The performance of the trained models on the DGkfold dataset is summarized in Fig. 3, which compares the Mean Absolute Error (MAE), Mean Squared Error (MSE), and Root Mean Squared Error (RMSE) for different algorithms on both the training and test sets. In terms of MAE, however, traditional linear models (PLSR and Ridge) and bagging-based ensemble models (ET and RF) show stable but limited accuracy for data governed by nonlinear kinetics. For instance, the MAE for Random Forest (RF) shows a noticeable deviation in Fig. 3(e). In contrast, ensemble models built upon gradient-boosting frameworks (namely CatBoost, LightGBM, GBDT, and XGBoost) consistently achieve superior predictive accuracy across most test scenarios. The highly consistent performance between the training and test sets not only verifies the models' generalization capability but also reflects effective overfitting control inherent in their design.

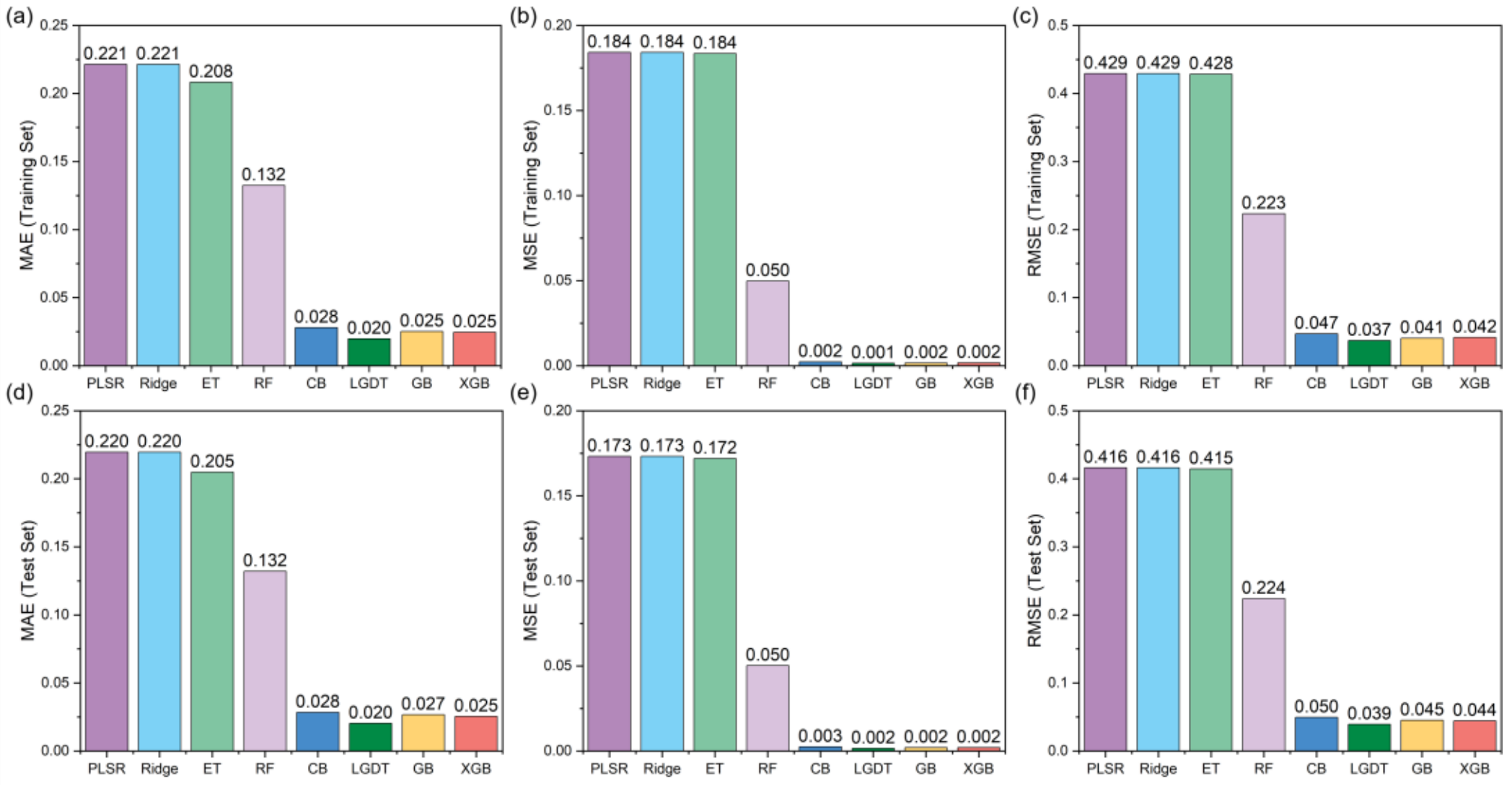

**Fig. 3.** Performance comparison of eight machine learning models, evaluated based on Mean

Absolute Error (MAE), Mean Squared Error (MSE), and Root Mean Squared Error (RMSE). (a-c) represent the errors on the training set, while (d-f) represent the errors on the test set.

A systematic evaluation of the goodness-of-fit and generalization ability is presented in Fig. S1(a–b). The coefficient of determination ($R^2$) exceeds 0.85 for all models on both training and test sets, indicating that the selected algorithms possess significant explanatory power for the reaction rate constant dataset. The gradient-boosting-based ensemble models (CatBoost, LightGBM, GBDT, and XGBoost) achieve consistently high $R^2$ values with $R^2 > 0.98$ on both datasets. The close agreement between training and test $R^2$ values ($\Delta R^2 < 0.02$) highlights these algorithms' capacity to capture underlying data patterns while effectively controlling overfitting. In contrast, the $R^2$ values for Ridge and PLSR decrease by 0.12–0.15 on the test set. Combined with their prediction bias standard deviation being 2.3–3.1 times larger than that of the ensemble models (Fig. S1(c–d)), this reveals the structural limitations of linear models in capturing the complex nonlinear kinetics involved. Integrating error distribution with $R^2$ analysis demonstrates that the ensemble learning framework, through adaptive feature weighting and residual iteration optimization, achieves precise modeling of the multi-dimensional coupling in reaction kinetic systems while maintaining prediction stability.

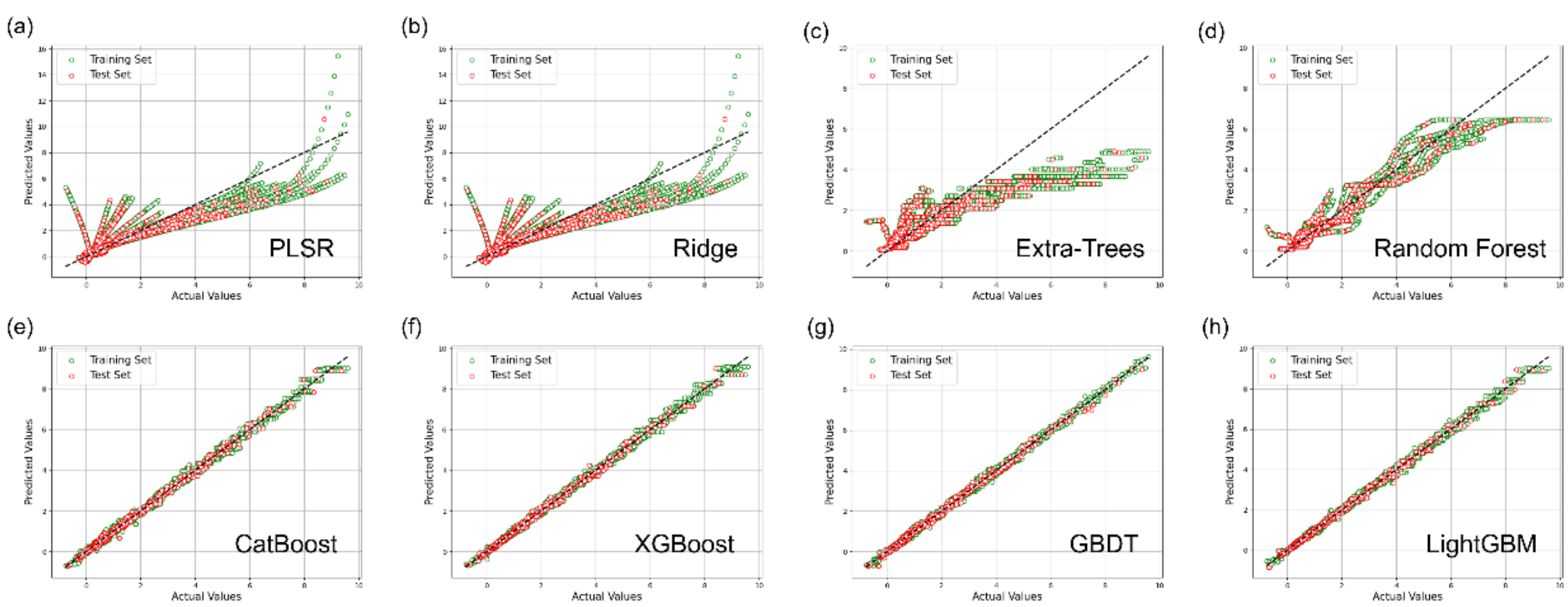

**Fig. 4.** Comparison of predicted values versus actual values for eight machine learning models.

Fig. 4 displays scatter plots of predicted versus actual values for the eight machine learning algorithms on both training and test sets. For Partial Least Squares Regression (PLSR, Fig. 4(a)) and Ridge Regression (Fig. 4(b)), although a high linear fitting quality is observed on the training set, the test set predictions (red) show significantly scattered distributions relative to the training set (blue), indicating overfitting risks likely arising from the linear methods' inability to represent the complex nonlinear relationships inherent in quantum tunneling effects. Extra-Trees (Fig. 4(c)) and Random Forest (Fig. 4(d)) perform well in the low-rate region, but in the high reaction-rate regime, RF predictions appear more clustered due to the averaging effect across multiple decision trees. In contrast, the gradient-boosting-based models, including CatBoost (CB, Fig. 4(e)), XGBoost (XB, Fig. 4(f)), Gradient Boosting Decision Trees (GBDT, Fig. 4(g)), and LightGBM (LG, Fig. 4(h)) demonstrate markedly superior performance: test set predictions align closely along the diagonal, with considerably reduced dispersion compared to traditional methods.

## 2. Validation of Model Generalization Capability

To systematically evaluate the generalization capability of the machine learning models to unseen reaction systems, a rigorous validation was performed using the leave-one-out cross-validation (LOOCV) dataset, DGLOO. The performance was assessed using root mean squared error (RMSE) and deviation distribution analysis across the eight algorithms (Fig. 5).

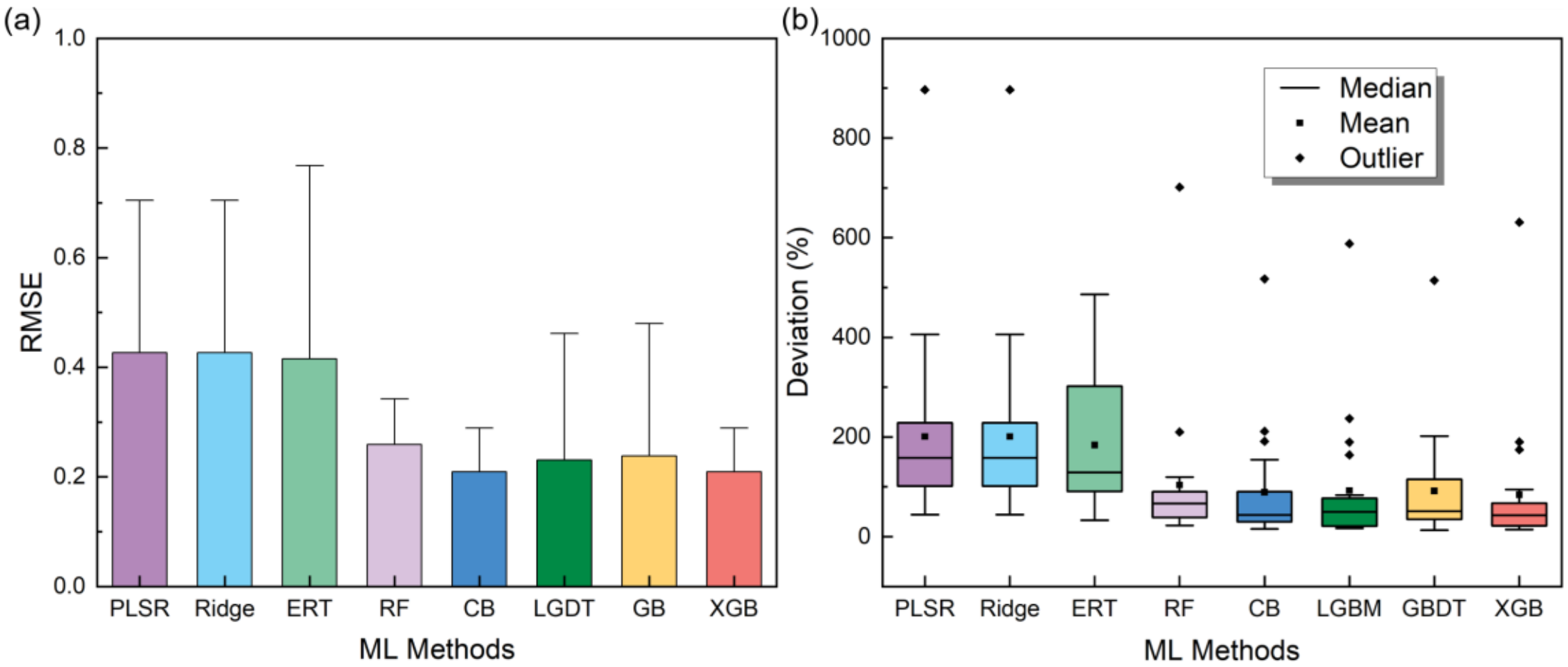

**Fig. 5.** Comparison of generalization performance for eight machine learning algorithms on a leave-one-out cross-validation dataset, evaluated based on Root Mean Squared Error (RMSE) and Deviation metrics.

As shown in Fig. 5(a), the traditional algorithms (PLSR, Ridge, and Extra-Trees) exhibited significantly higher average RMSE values (0.42, 0.42, and 0.40, respectively) compared to the ensemble methods (XGBoost: 0.21, CatBoost: 0.21). While Random Forest (RF) performed best among the traditional methods (RMSE = 0.30), a clear performance gap remained relative to the gradient-boosting frameworks. This disparity can be attributed to the superior capacity of ensemble algorithms to model high-dimensional, nonlinear couplings. Further analysis of the deviation distribution (Fig. 5(b)) reveals that XGBoost possesses the smallest interquartile range in its boxplot, indicating its consistently superior generalization across diverse reaction pathways compared to other algorithms.

## 3. Feature Importance Analysis and Physical Interpretation

A comparative analysis of feature contributions across multiple models reveals algorithm-specific differences in interpretability regarding the key drivers for reaction rate constant prediction, as illustrated in Fig. S2. Within tree-based ensemble models, temperature (T) and the KIE exhibit significant global influence. Extra-Trees (ET) and Random Forest (RF) assign higher relative importance to T and the reaction rate (k), respectively, while CatBoost (CB), GBDT (GB), and XGBoost (XGB) emphasize the central role of KIE. The average weight of PES asymmetry (η) in CB reaches ~30%, second only to KIE and significantly higher than its ~15% level in GB/XGB models, which likely reflects CatBoost's symmetric tree architecture, which handles ordered categorical features differently from conventional gradient boosting. Although the distribution of feature weights varies algorithmically, KIE consistently holds a dominant cumulative weight across the six mainstream ensemble models, with other factors (η, k, and T) also playing secondary but measurable roles.

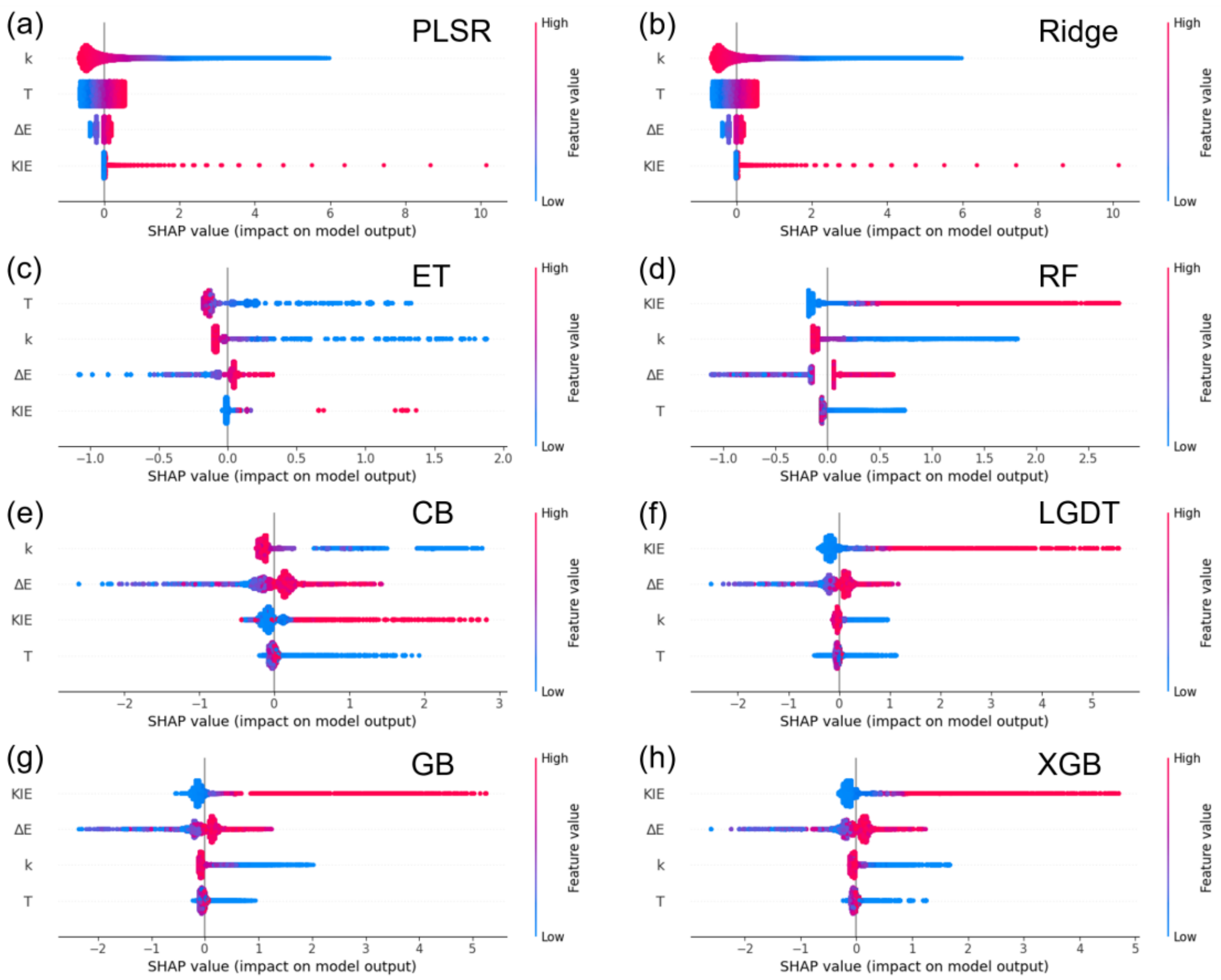

**Fig. 6.** Overview diagram of SHAP analysis for eight machine learning algorithms.

To complement the feature importance analysis, SHAP (SHapley Additive exPlanations) analysis offers an approach that additionally provides directional feature importance. Rooted in the concept of Shapley values from cooperative game theory, SHAP-based interpretability analysis (Fig. 6) elucidates the differential feature dependence mechanisms across algorithms for predicting the tunneling strength metric κ. Based on the performance evaluation on the DGLOO validation set, we focus on the results from the XGBoost algorithm. Fig. 6 (f) reveals that κ exhibits the strongest correlation with KIE, followed by PES asymmetry, reaction rate, and temperature. The color gradient from blue to red corresponds to increasing KIE values, and the associated SHAP values also generally increase, suggesting an overall positive correlation. However, the data density is higher at lower KIE values and more uniformly distributed at higher KIE values, indicative of a complex, nonlinear relationship.

## 4. Tunneling Phase Diagram: Revealing the Nonlinear KIE–κ Relationship

Building upon the discussions of algorithmic reliability and generalization, we used the best-performing XGBoost algorithm to construct a quantitative KIE–κ mapping, as shown in Fig. 7.

Having established that KIE alone is insufficient to predict κ, we constructed a tunneling phase diagram based on the XGBoost model to quantitatively map the KIE–κ relationship. Our results demonstrate that relying solely on KIE can lead to misinterpretation. Specifically, in the low-temperature regime (100–200 K, Fig. 7(a)-(b)), the strong tunneling region exhibits a broad distribution across KIE values. The high dispersion reflects the sensitivity of the isotopic effect to barrier morphology (e.g., asymmetry η) and quantum tunneling pathways. As temperature increases (300–800 K, Fig. 7(c)-(h)), the KIE distribution gradually converges, and its numerical decay coincides with the decreasing trend of κ, indicating the progressive dominance of classical thermodynamic mechanisms over quantum tunneling.

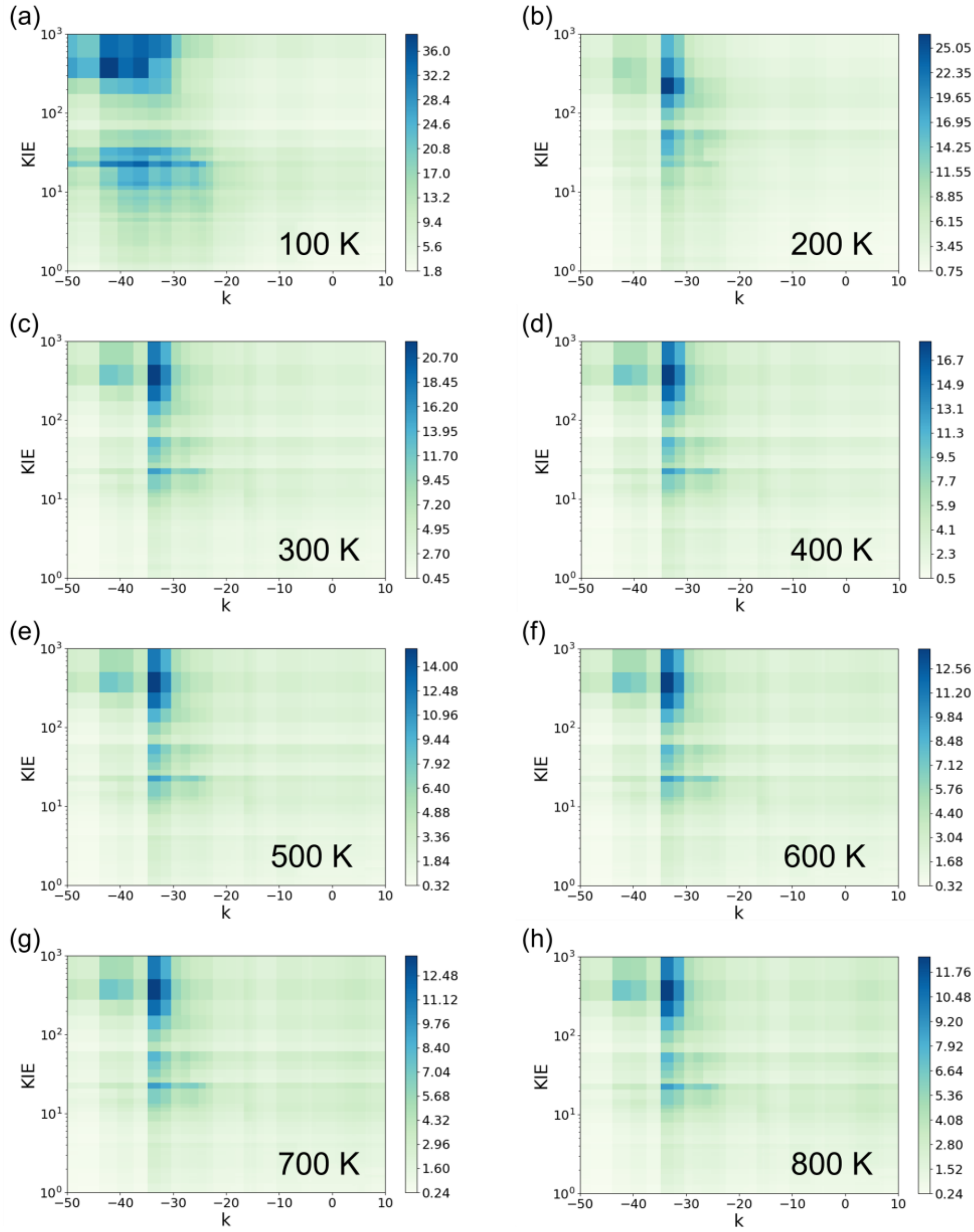

**Fig. 7.** Relationship between the KIE and the tunneling correction factor (κ). (a)-(h) represent the results obtained at different temperatures.

A key finding emerges in the intermediate temperature range (300–600 K, Fig. 7(c)–(f)): although KIE values remain relatively high, the rapid decline in κ reveals a crossover between classical thermal effects and quantum tunneling. When the quantum reaction rate exceeds ~$10^{-5}$ $s^{-1}$ (right regions of Fig. 7(e)–(f)), the combined influence of barrier width and asymmetry significantly attenuates the tunneling probability, giving rise to an anomalous regime characterized by high KIE–low κ. This regime represents a direct counterexample to the conventional assumption that larger KIE implies stronger tunneling, and it underscores the central message of this work: tunneling strength cannot be reliably inferred from KIE alone.

The regional characteristics of the tunneling phase diagram guide both experimental design and theoretical modeling. In the low-temperature, low-rate region (Fig. 7(a)-(b), $T < 300$ K, $k < 10^{-5}$ $s^{-1}$), κ values are generally above 2.0, corresponding to a tunneling-dominated regime where symmetric barrier morphology promotes quantum effects of the hydrogen nucleus. Conversely, in the high-temperature, high-rate region (Fig. 7(g)-(h), $T > 600$ K, $k > 10^0$ $s^{-1}$), κ values approach 1.0, indicating the diminishing contribution of quantum tunneling relative to classical over-the-barrier processes. The intermediate transition region (Fig. 7(c)-(f), $300 K < T < 600 K$) is particularly noteworthy, exhibiting a significant gradient in κ and a clear KIE–κ decoupling phenomenon. Therefore, the practical assessment of tunneling strength (κ) necessitates a multidimensional criterion incorporating temperature, reaction rate, and barrier descriptors (e.g., η), rather than relying on a single KIE threshold.

## Conclusion:

The tunneling phase diagram introduced in this work provides a quantitative framework for decoding the nonlinear relationship between the kinetic isotope effect and the intrinsic tunneling correction factor κ. By integrating machine learning with first-principles rate calculations, ensemble models capture this multidimensional mapping with high accuracy ($R^2 > 0.98$, RMSE = 0.21). SHAP analysis confirms that the KIE–κ relationship is strongly modulated by temperature and barrier asymmetry, underscoring the necessity of a multidimensional approach beyond single-descriptor heuristics. It maps experimentally accessible variables—KIE, temperature, and reaction rate—to κ, revealing distinct dynamical regimes and explicitly identifying regions where KIE and κ decouple. This provides a practical tool for estimating tunneling contributions from routine kinetic measurements. By shifting from a single-descriptor heuristic to a multidimensional, physically grounded approach, this work establishes a new framework for the quantitative assessment of quantum tunneling in proton and hydrogen-atom transfer reactions.

## Conflict of Interest:

The authors declare no competing financial interests.

## Acknowledgments:

This work was supported by the Science and Technology Development Program of Jilin Province of China (20250102014JC) and the National Key Research and Development Program of China (No. 2024YFA1409900). Z. Wang also acknowledges the assistance of the High-Performance Computing Center of Jilin University and National Supercomputing Center in Shanghai.

## Supporting Information:

Supporting Information is available: hyperparameter optimization results (Table S1-S4), additional performance evaluation (Fig. S1), and feature importance analysis (Fig. S2).